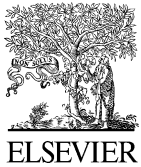

GHGT-12

# Uncertainty quantification for $CO_2$ sequestration and enhanced oil recovery

Zhenxue Dai[a]*, Hari Viswanathan[a], Julianna Fessenden-Rahn[a], Richard Middleton[a], Feng Pan[b], Wei Jia[b], Si-Yong Lee[c], Brian McPherson[b], William Ampomah[d] and Reid Grigg[d]

[a]Earth and Environmental Sciences Division, Los Alamos National Laboratory, Los Alamos, NM 87545, USA
[b]Energy and Geoscience Institute, The University of Utah, Salt Lake City, UT 84108, USA
[c]Schlumberger Carbon Services, Cambridge, MA 02139, USA
[d]Petroleum Recovery Research Center, New Mexico Tech, Socorro, NM 87801

## Abstract

This study develops a statistical method to perform uncertainty quantification for understanding $CO_2$ storage potential within an enhanced oil recovery (EOR) environment at the Farnsworth Unit of the Anadarko Basin in northern Texas. A set of geostatistical-based Monte Carlo simulations of $CO_2$-oil-water flow and reactive transport in the Morrow formation are conducted for global sensitivity and statistical analysis of the major uncertainty metrics: net $CO_2$ injection, cumulative oil production, cumulative gas ($CH_4$) production, and net water injection. A global sensitivity and response surface analysis indicates that reservoir permeability, porosity, and thickness are the major intrinsic reservoir parameters that control net $CO_2$ injection/storage and oil/gas recovery rates. The well spacing and the initial water saturation also have large impact on the oil/gas recovery rates. Further, this study has revealed key insights into the potential behavior and the operational parameters of $CO_2$ sequestration at $CO_2$-EOR sites, including the impact of reservoir characterization uncertainty; understanding this uncertainty is critical in terms of economic decision making and the cost-effectiveness of $CO_2$ storage through EOR.

© 2013 The Authors. Published by Elsevier Ltd.
Selection and peer-review under responsibility of GHGT.

* Corresponding author. Tel.: +01-505-665-6387; fax: +01-505-665-8737.
E-mail address: daiz@lanl.gov

## 1. Introduction

Carbon sequestration with enhanced oil recovery ($CO_2$-EOR) is a promising technology for emissions management because of its low cost in the absence of emissions policies that include incentives for carbon capture and storage [1]. And, the demand for $CO_2$ will only increase as more domestic oil is produced using $CO_2$-EOR. Currently, $CO_2$-EOR provides only about 5% percent of the total U.S. crude oil production [2]. With advancement of sequestration technology, infrastructure, and regulations, more anthropogenic $CO_2$ sources will be available for $CO_2$-EOR in the next few decades. Most ongoing $CO_2$-EOR uses water-alternating-gas (WAG) to control $CO_2$ mobility and $CO_2$ flood conformance (flood front uniformity) and to tackle clogging and scale issues associated with oil reservoirs [3]. While WAG can be very effective, a few operational and technical difficulties for commercial $CO_2$-EOR still remain: (1) highly-heterogeneous reservoirs are difficult to characterize and it is even more difficult to quantify the impact of reservoir heterogeneity on $CO_2$ injectivity and oil/gas production; (2) guidelines for determining well spacing are not robust or general; and (3) time ratios of WAG injection are difficult to evaluate quantitatively [4-16].

These issues have served as a motivation for this study. The goal of this study is to develop a statistical method to quantify uncertainty in $CO_2$ sequestration and enhanced oil recovery for depleted oil reservoirs. This method consists of a multi-phase reservoir simulator coupled with geologic and statistical models to characterize reservoir heterogeneity and to sample associated uncertain parameters. A set of Monte Carlo simulations of $CO_2$-oil-water flow and reactive transport is conducted for the Morrow reservoir in the Farnsworth Unit of the Anadarko Basin in northern Texas, followed by a global sensitivity, response surface, and general statistical analysis. For quantitative evaluation of the operational and technical uncertainty of the $CO_2$-EOR systems, we defined a set of uncertainty metrics to post-process the Monte Carlo simulation results for statistical analysis. The metrics include: net $CO_2$ injection, cumulative oil production, cumulative $CH_4$ production, and net water injection. The two-dimensional and three-dimensional response surfaces are developed based on the known parameter ranges and distributions in the Morrow reservoir. The developed response surfaces also facilitate a statistical analysis for estimating the mean, and confidence intervals of the uncertainty metrics.

## 2. Characterization of reservoir heterogeneity and geostatistical modeling

The Farnsworth Unit is located in the Anadarko Basin of northern Texas. Southwest Partnership and Chaparral selected the major oil/gas reservoir, the Morrow formation, as its primary test reservoir to evaluate long-term storage of $CO_2$ [17]. Previous studies in nearby oil fields provide prior information about the distributions of regional Morrow reservoir parameters, such as the depth, thickness, permeability, and porosity. The regional Morrow reservoir mainly consists of incised valley-fill sandstones of the Lower Pennsylvanian that extend into eastern Colorado and western Kansas [18-21]. The regional reservoir in the Anadarko Basin has produced more than 100 million barrels of oil and 14.2 billion $m^3$ of gas. Figure 1 shows the positive correlation between the measured permeability and porosity in the regional Morrow reservoir (modified from Bowen [22]). The blue dots represent permeability and porosity data collected from medium to coarse grained sandstone; the pink dots represent the data from fine grained sandstones with mud drapes; the yellow triangle symbols are data from cemented sandstone; the cross symbols are data from fine-grained cross-bedded sandstone; and the star symbols are data from transgressive lag. The data points within the green circle represent permeability and porosity distributions at the Farnsworth site.

Table 1: Statistics of the measured Morrow reservoir parameters (54 sampled points)

| Statistics | Sample Depth (m) | Grain Density (g/cm³) | Porosity (%) | Permeability (mD) | Permeability (logm²) | Water Saturation (%) | Oil Saturation (%) |
|---|---|---|---|---|---|---|---|
| Minimum | 2337.24 | 2.63 | 5.49 | 0.20 | -15.71 | 10.68 | 8.69 |
| Maximum | 2348.73 | 2.92 | 22.69 | 783.50 | -12.11 | 58.03 | 30.69 |
| Mean | 2343.02 | 2.67 | 16.78 | 69.21 | -13.70 | 21.94 | 21.31 |
| Standard Deviation | 3.47 | 0.05 | 3.72 | 130.93 | 0.79 | 8.31 | 4.50 |

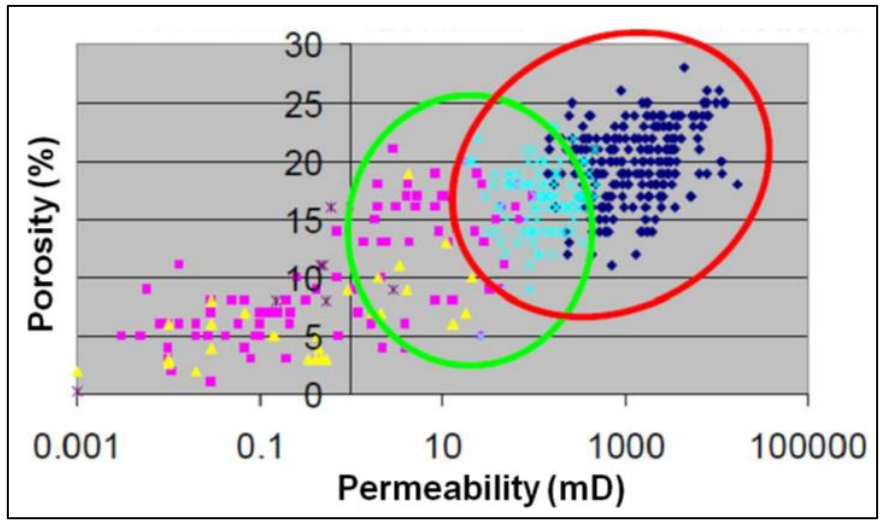

Figure 1: The positive correlation between the measured permeability and porosity in the Morrow formation (Modified from Bowen [22])

The Morrow formation at the Farnsworth site is located at depths around 2330 m and typical thicknesses between 5 and 50 m [18-21]. Recently, more measurements of the Morrow reservoir parameters were collected at the Farnsworth site by Southwest Partnership and Chaparral [23]. A summary of the measured parameters is listed in Table 1. The permeability in this site is estimated between 0.2 and 783.5 milli-Darcy (mD) and the porosity is between 0.05 and 0.23. The computed Pearson's correlation coefficient between porosity and log permeability is about 0.8, which means these two parameters are highly correlated. Note that the Pearson's coefficient is a measure of the strength and direction of the linear relationship between two parameters that is defined as the sample covariance of the variables divided by the product of their sample standard deviations. According to Bernabe et al. [24] and Deng et al. [25], the relationship between permeability and porosity is expressed as:

$$k = a \, \phi^b, \qquad (1)$$

where $k$ is permeability ($m^2$), $\phi$ is porosity, $a$ and $b$ are constants. By using the permeability and porosity data collected in this site, we estimated the two constants: $a = 345.2$ and $b = 4.1$. The measured and computed permeability vs porosity is plotted in Figure 2.

Based on the prior parameter information in the regional Morrow reservoir and the statistics of recently-measured parameters, we summarize the ranges and distributions of the uncertain parameters for simulating the heterogeneity of the Morrow reservoir in the Farnsworth site in Table 2. Having limited existing spatial-based permeability data in this site, we assume that the horizontal and vertical integral scales are 500 m and 50 m in the reservoir, respectively. The permeability anisotropy factor (or ratio of vertical and horizontal permeability) is assumed to be 0.1. The relative permeability functions for $CO_2$-oil-water multiphase flow simulations were calculated on the base of Stone's approach to define the related coefficients [26-29]. Table 2 also lists the range of time ratio of WAG for alternatively injecting $CO_2$ gas and water within each time period or cycle (such as 10 days or 20 days). This injection time ratio is calculated by dividing $CO_2$ injection time (days) by water injection time (days) in each time period.

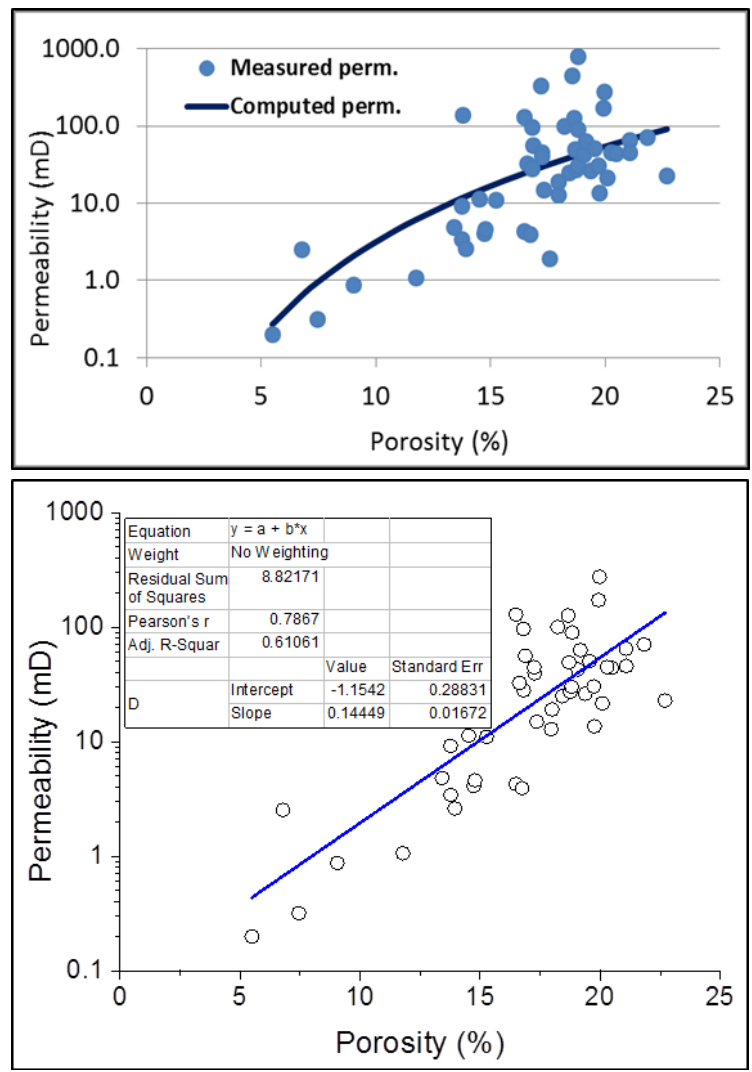

Figure 2: The positive correlation between the measured permeability and porosity in the Morrow formation

## 3. Integrated Monte Carlo simulations

A set of integrated Monte Carlo simulations of $CO_2$-oil-water flow and transport in the reservoir is developed by coupling the uncertainty quantification tool PSUADE [30], the Los Alamos developed geostatistical modeling tool GEOST [31-33] modified from the Geostatistical Software Library [34], and the multi-phase reservoir simulator SENSOR [27]. PSUADE is used to sample 1000 realizations of the uncertain parameters with Latin Hypercube Sampling, to conduct global sensitivity analysis of the output variables in relation to the uncertain parameters, and to derive response surfaces or reduced order models (ROMs) for understanding the relationships of the input parameters and the output variables. GEOST is used to analyze the existing permeability data and to generate heterogeneous permeability distributions for the reservoir with a sequential Gauss method. The reservoir porosity is computed by the correlation equation (1) with the generated permeability data. The reservoir simulator SENSOR is used to model $CO_2$-oil-water flow and transport in the reservoir for each generated heterogeneous model under a Monte Carlo simulation framework.

Table 2: Uncertain parameters and objective functions for the Farnsworth site

|  | Variables names | Min. | Max. | Mean |
|---|---|---|---|---|
| **Reservoir (Morrow) Parameters** | Thickness(m) | 5.0 | 50 | 27.5 |
|  | Permeability(mD) | 0.2 | 783.5 | 0.069 |
|  | Porosity | 0.05 | 0.23 | correlated |
|  | Initial water saturation | 0.11 | 0.58 | 0.22 |
|  | Initial oil saturation | 0.09 | 0.31 | 0.21 |
|  | Well spacing (km) | 0.1 | 0.5 | / |
|  | Time ratio of WAG | 0.0 | 10 | / |
| **Uncertainty metrics** | Net $CO_2$ injection(Mton) |  |  |  |
|  | Oil production(MMbbl) |  |  |  |
|  | Gas production($m^3$) |  |  |  |
|  | Net water injection(Mton) |  |  |  |

A five-spot pattern (Figure 3) is selected for this study, where the production well is located in the center surrounded by four injection wells at the corners of the pattern with a well spacing (or injection distance). The heterogeneity in the model area is assumed to be symmetric in all four quadrants of the EOR pattern. This way, only one quarter of the five-spot pattern is required in the model with one injection well and one fourth of the production well, which implies that all of the six boundaries (top, bottom, front, back, left and right) are fixed as no-flow. The numerical model sizes and grid numbers are automatically calculated based on the sampled well spacing and reservoir thickness and a uniform mesh is used for each realization. The reservoir simulations using SENSOR start from the sampled initial water and oil saturations at a steady state and then simulate $CO_2$ injection and oil/$CH_4$ production at the Farnsworth site over 5 years for the 1000 realizations. For each realization a post-processing step is conducted to compute statistics on four uncertain metrics 1) net $CO_2$ injection which is the difference of injection and production $CO_2$ rates in 5 years (CO2inj), 2) cumulative oil production (oilprod), 3) cumulative $CH_4$ production (CH4prod), and 4) net water injection which is the difference of injection and production water rates in 5 years (H2Oinj).

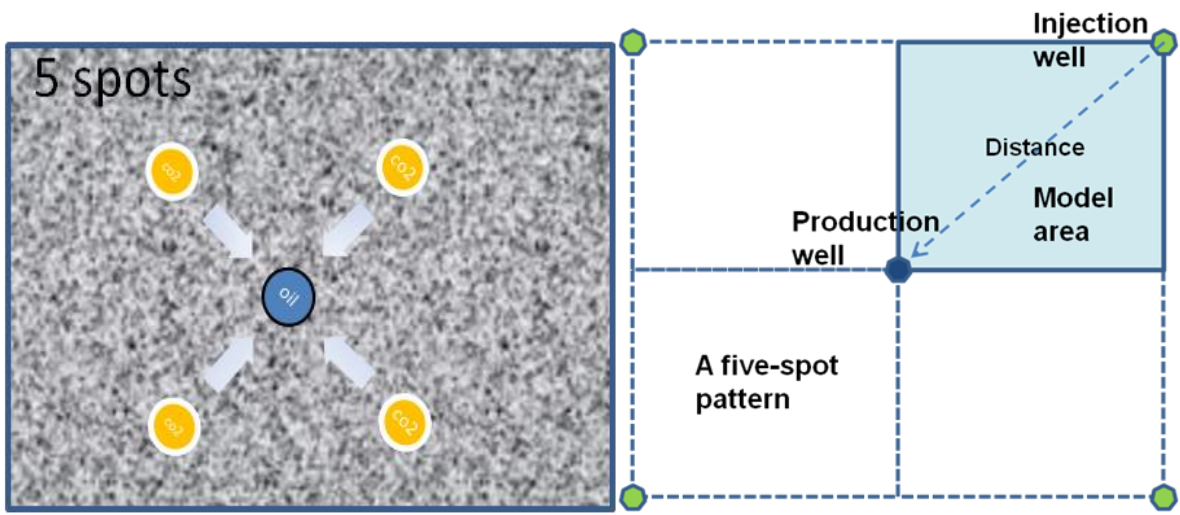

Figure 3: A five-spot $CO_2$-EOR pattern and the model permeability distributions for optimizing the injection well distance and time ratio of WAG.

## 4. Global sensitivity analysis with MARS method

In order to determine the key flow and transport parameters driving $CO_2$–oil/gas-water migration behavior in the reservoir, global sensitivity analysis techniques were used for investigating input-output sensitivities over the entire distributions of the uncertain parameters [35-36]. The *multivariate adaptive regression spline* (MARS) method was used to quantify the impact of uncertainty and sensitivity of the input parameters.

By using the Monte Carlo simulation results as the input for PSUADE [30], we conduct global sensitivity analysis with the MARS method for the four risk uncertainty metrics. The results plotted in Figure 4 show that different metrics are sensitive to different parameters. The net $CO_2$ injection is mainly controlled by the time ratio of WAG, reservoir permeability, thickness, and porosity (Figure 4A). The oil production is most sensitive to the reservoir porosity, well spacing, thickness, and initial water saturation (Figure 4B). The $CH_4$ (gas) production is most sensitive to the initial water saturation, well spacing, reservoir thickness, porosity, and permeability (Figure 4C). Further correlation analysis indicates that the oil and gas ($CH_4$) production rates are negatively correlated to the initial water saturation. Finally, Figure 4D shows that the net water injection is most sensitive to the reservoir permeability, time ratio of WAG, well spacing and thickness.

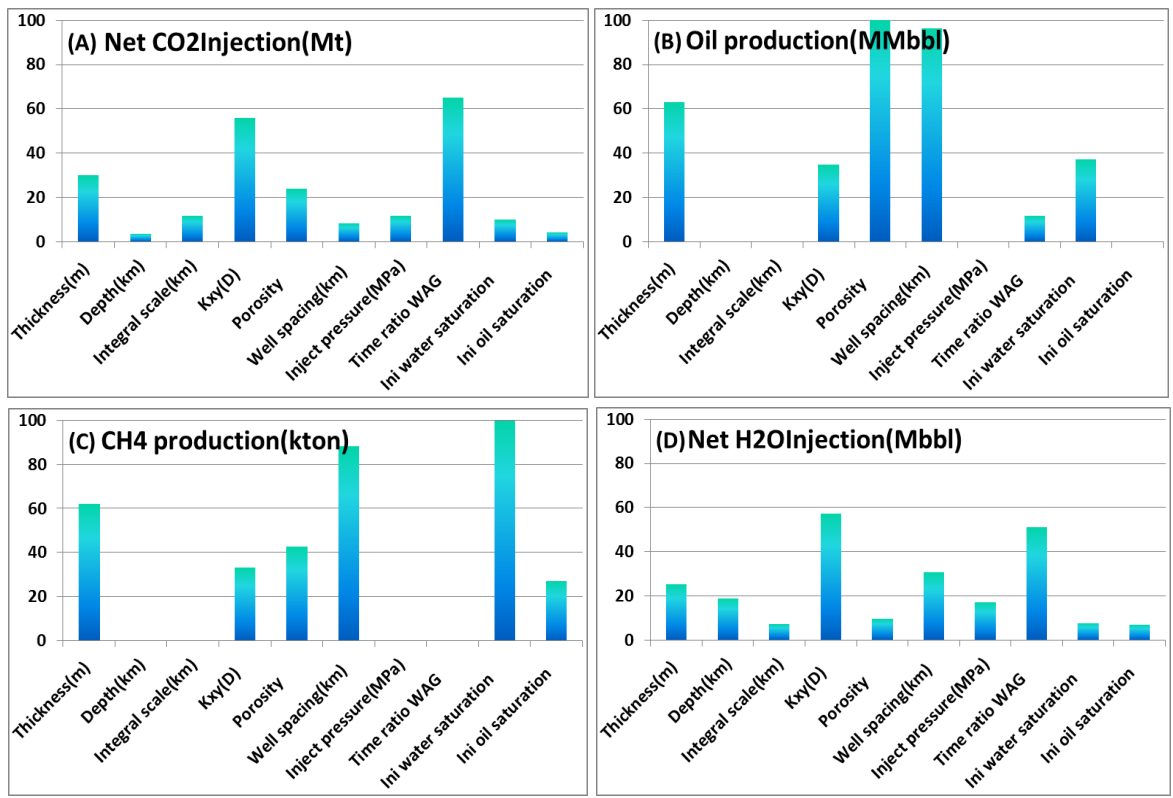

Figure 4: Global sensitivity analysis of the risk metrics to the uncertain parameters using MARS method [30]

## 5. Response surface analysis

Response surface analysis is an application of statistical and mathematical techniques useful for developing and reducing the orders of the process models. Using the post-processing results of the 1000 Monte Carlo simulations, we conducted a response surface analysis of the uncertainty metrics by using the MARS approach with bootstrap aggregating (bagging) [30]. The fitting results of the regression to generate the response surfaces are presented in Figure 5. The corresponding $R^2$ of these three response surfaces are larger than 0.95, respectively, which means the generated response surfaces can represent the process models very well. Figure 6 shows the two- and three-dimensional plots of the MARS response surfaces for $CO_2$ injection and oil and gas ($CH_4$) production in relation to the most sensitive parameters. The net $CO_2$ injection rates and oil/gas production rates are positively correlated to reservoir porosity, permeability and thickness. Note that for the 1000 Monte Carlo runs the $CO_2$ injection pressures are assumed to be 70% of the hydrostatic pressures at the reservoir tops, which causes the $CO_2$ injection rates and oil/gas production rates to be positively correlated to the reservoir depths. The developed response surfaces will be utilized in the risk assessment framework, $CO_2$-PENS [37-38], for evaluating $CO_2$, oil, gas ($CH_4$), and brine water interactions under the $CO_2$-EOR environment at the Farnsworth site.

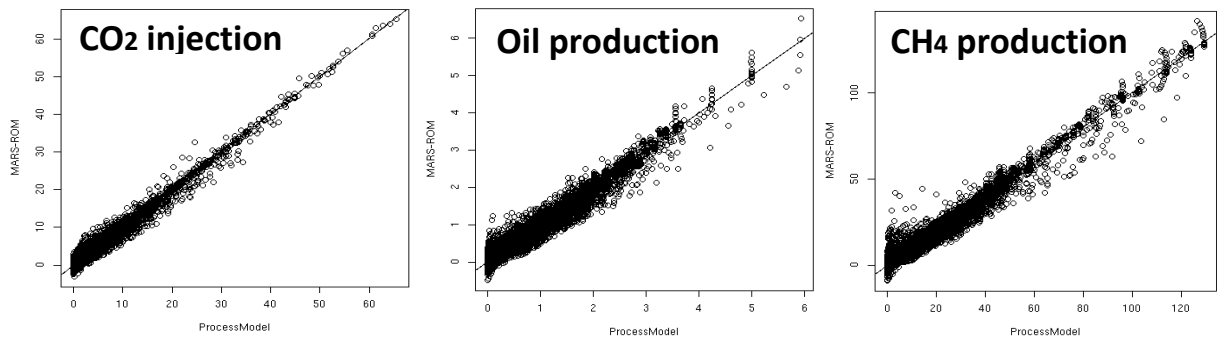

Figure 5: The fitting results of the regression to generate the response surfaces of $CO_2$ injection, oil and gas ($CH_4$) production.

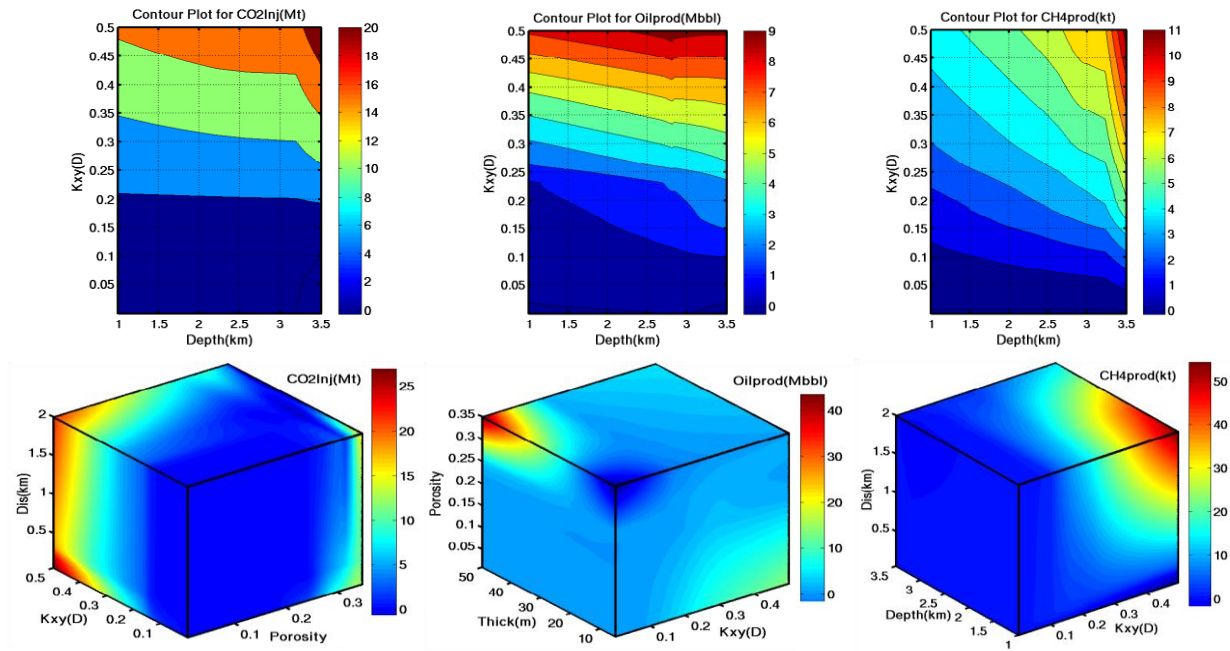

Figure 6: Two- and three-dimensional response surfaces generated with regression methods to show the relationships of the major input parameters and the risk metrics.

## 6. Summary and conclusions

This study captures the complex multi-phase flow and transport processes of $CO_2$-oil/gas-water in the reservoir and generates the computationally efficient response surfaces. These response surfaces are in relation to the uncertain parameters in a reduced order form. The global sensitivity results indicate that the reservoir thickness, permeability, and porosity are the key parameters to control the $CO_2$/water injection and oil/$CH_4$ production rates. The distance between the injection and production wells (well spacing) and the initial water saturation also have a large impact on the oil/$CH_4$ production.

The response surface analysis shows that net $CO_2$ injection rate increases with the increasing reservoir thickness, permeability, porosity and well spacing. The oil/$CH_4$ production rates are positively correlated to reservoir permeability, porosity and thickness, but negatively correlated to the initial water saturation.

Our next step will replace the current calculations with more complex STOMP models [39] to estimate the net $CO_2$ and water injection, oil/$CH_4$ production rates, and the optimal distance between the injection and production wells. The developed response surfaces will be utilized in the risk assessment framework, $CO_2$-PENS [37-38], to account for $CO_2$, oil, gas ($CH_4$), and brine water interactions and to understand the impact of the reservoir heterogeneity uncertainty on the critical economic decision making and the cost-effectiveness of $CO_2$ storage through EOR at this site.

### Acknowledgements

This work is part of the Phase III of the SWP $CO_2$-EOR/Storage Project that is supported by the U.S. Department of Energy and managed by the National Energy Technology Laboratory. We gratefully acknowledge the assistance of Brian Coats of Coats Engineering, Inc. for providing the multi-phase flow and transport modeling code and discussing with us the multi-phase model setup.